\documentstyle[aps,prb,twocolumn,epsf]{revtex}

\begin{document}

\title{Thermopower peak in phase transition region of (1-x)La$_{2/3}$Ca$_{1/3}$MnO$_{3}$/xYSZ}

\author{Wei Liu, Jun Zhao, Chinping Chen\cite{chen},  and Shousheng Yan}

\address{
Department of Physics and State Key Lab for Mesoscopic Physics,
Peking University, Beijing 100871, P.\,R.\,China }

\author{Zhengcai Xia, Sheng Liu}

\address{Department of Physics, Huazhong University of Science and
Technology, \\ Wuhan 430074, P.\,R.\,China }

\maketitle

\begin{abstract}
The thermoelectric power (TEP) and the electrical resistivity of
the intergranular magnetoresistance (IGMR) composite,
(1-x)La$_{2/3}$Ca$_{1/3}$MnO$_{3}$/xYSZ ( LCMO/YSZ ) with x = 0,
0.75\%, 1.25\%, 4.5\%, 13\% 15\% and 80\% of the yttria-stabalized
zirconia (YSZ), have been measured from 300 K down to 77 K.
Pronounced TEP peak appears during the phase transition for the
samples of x $>$ 0, while not observed for x = 0. We suggest that
this is due to the magnetic structure variation induced by the
lattice strain which is resulting from the LCMO/YSZ boundary
layers. The transition width in temperature derived from
$d\chi/dT$, with $\chi$ being the AC magnetic susceptibility,
supports this interpretation.
\end{abstract}

\pacs{To be added}

\section{Introduction}

The colossal magnetoresistance (CMR) material has been one of the
focused points in the research community due to its promising
potential in the applications and its complicated properties in
the basic physics research. Many models have been proposed in the
early days to discuss the observed phenomena, from the double
exchange model \cite{Zener51,Anderson55,Goodenough55,deGennes60}
for the ferromagnetic phase at low temperature (LT-FM phase) to
the bi-polaron model \cite{Alexandrov99a,Alexandrov99b} for the
paramagnetic phase at high temperature (HT-PM phase). However,
none of these alone can satisfactorily explain the complicated
behaviors occurring during the Metal/Insulator (MI) or the
Ferromagnetic/Paramagnetic (FM/PM) phase transition. Deeper
understanding into the physics of the CMR phase transition is
therefore necessary. Among the existing models, the percolation
theory of the FM/PM coexistence in the phase transition has
received much attention with many supporting evidences
\cite{Bastiaansen98,Freisem99,Merithew00}.

In some of the previous reports for the TEP measurements on the
manganites, a pronounced positive peak or an abrupt jump has been
observed during the phase transition
\cite{Jaime96,Jaime99,Fontcuberta96}. However, there have been
other reports on the samples of the same compositions without
exhibiting any peak or jump in the TEP during the phase transition
\cite{Hundley97,Mandal00}. In one of our previous
experiments\cite{Liu03}, the
(La$_{1-x}$Y$_{x}$)$_{2/3}$Ca$_{1/3}$MnO$_{3}$ samples with x = 0,
0.05, 0.15, 0.20 have been studied. There is a TEP peak showing up
during the phase transition for the high doping sample, x = 0.2,
while not observed for the other low doping ones. An intermediate
phase resulting from the magnetic structural inhomogeneity induced
by the lattice mismatch, which is a direct result of the random
distribution of different Perovskites A-site cations, has been
proposed to explain this phenomenon. By the similar mechanism, any
defect-induced mismatch or the strain propagated from the
interface layers in the lattice would introduce similar effect on
the magnetic structure. Hence a TEP jump would appear during the
phase transition. This behavior in TEP have been observed in the
thin film sample of single crystal La$_{2/3}$Ca$_{1/3}$MnO$_{3}$
\cite{Jaime96,Jaime99}, and in the samples of
La$_{0.7}$Ca$_{0.3}$MnO$_{3}$ by different heat treatment in the
oxygen environment \cite{Mahendiran96}. The underlying cause for
the TEP peaks to occur is perhaps other than an intrinsic one as
suggested with the charge-orbit coupling \cite{Nakamae01} or the
band narrowing \cite{Fontcuberta96} effects. Instead, the lattice
mismatch caused by the strain resulting from the substrate for the
former case and from the deficiency of the oxygen for the latter
one would, perhaps, provide a consistent picture for the
understanding.

In order to gain deeper insights into the properties of the CMR
manganites, especially during the phase transition, we have
carried out extensive measurements on the series of the samples,
the (1-x)La$_{2/3}$Ca$_{1/3}$MnO$_{3}$/xYSZ compound with x = 0,
0.75\%, 1.25\%, 4.5\%, 13\%, 15\%, and 80\%. These include the
temperature dependent resistivity ($\rho-T$), the TEP, and the AC
susceptibility measurements. The correlation of the TEP peak with
the YSZ composite ratio, x, has been studied. The lattice strain
induced by the boundary layers of the YSZ insulator intermixed
with the LCMO phase provides a satisfactory basis for the
understanding of the TEP peak to appear during the phase
transition. In addition, the structure effect of the two-component
composite on the electrical transport properties and the phase
transition temperature $T_{C}$ has been explained within the
framework of the percolation theory.

\section{Sample preparation}

The LCMO/YSZ samples were fabricated by double-staged production
procedure as reported previously \cite{Yuan02}. Nanometric powders
of the LCMO was first prepared by the sol-gel method and then
sintered at 1300$\ ^{o}C$ for 10 hours to obtain powdered crystals
of grain size $\sim$ 20 $\mu$m. Then, it was mixed thoroughly with
the YSZ powders of grain size $\sim$ 2 $\mu$m for the heat
treatment in the ambient air environment at 1350$\ ^{o}C$ for
another 10 hours. The crystal structure was then characterized by
the X-ray diffraction (XRD) analysis using the Cu K$_{\alpha}$
source. The YSZ phase is identified for the samples of  x $> 5\%$,
and the lattice parameters of the LCMO phase remain unchanged
within the level of 0.001 $\AA$ for all of the samples. This
evidence indicates that the LCMO and the YSZ phases exist
independently to form solid intermixture, without the
inter-diffusion of the ions from the YSZ phase into the LCMO phase
during the heat treatment. The results by the SEM images provide
further evidence that the YSZ phase exists separately from that of
the LCMO \cite{Yuan02}.

\section{Experiment}

The zero-field temperature dependent resistivity ($\rho-T$)
measurements have been carried out by the standard 4-probe method.
Typical $\rho-T$ curves of the CMR materials are obtained and
presented in Fig.~\ref{RT-peak} for all of the samples except for
the one of x = 80\%, which has the insulator property with the
resistance $> 10^{12}\,\Omega$. The temperature and the
resistivity of the MI transition peak have been plotted against
the mixing ratio, x, in the inset of Fig.~\ref{RT-peak}. The
zero-field $\rho-T$ property from the present work is consistent
with the result reported in the previous experiment \cite{Yuan02}.
The lowest transition temperature occurs at x$_{m}$ = 4.5\% with
the corresponding peak resistivity having the largest value, see
the inset of Fig.~\ref{RT-peak}. At the low mixing ratio with x
$<$ x$_{m}$, the distribution of the YSZ forms insulation layers
encompassing the LCMO clusters completely. This has been confirmed
by the direct SEM observation. At this structure, the mechanism of
the electrical conductance between the LCMO phase separated by the
YSZ phase is realized by the tunnelling conduction through the
insulating YSZ layer. With the x increasing from 0 to x$_{m}$, the
size of the LCMO cluster decreases with the inter-granular effect
going up\cite{Yuan02}. Accordingly, the peak resistivity,
$\rho_{peak}$, increases and the peak temperature, $T_{\rho}$,
goes down, at the increasing x, see the inset of
Fig.~\ref{RT-peak}. On the other hand, for the high x samples at x
$>$ x$_{m}$, the insulating YSZ phase tends to form cluster-like
structure and the LCMO phase would constitute a 3-dimensional
percolation connection without being surrounded completely by the
YSZ phase. This makes the conduction behavior of the high x
samples tends to resemble the one of x = 0 at the LT-FM phase. For
the sample of x = 80 \%, which exceeds the threshold density by
the percolation theory, the conducting passage is disconnected
within the sample by the YSZ insulator and the insulation state is
resulted.

The temperature dependent TEP measurement ($S-T$ measurement) has
been conducted by the differential DC method from 320 K down to 77
K. The results are shown in Fig.~\ref{TEP}. For the particular
sample of x = 0, which is La$_{2/3}$Ca$_{1/3}$MnO$_3$, the TEP
shows a smooth transition from the low temperature metallic
behavior to the high temperature $1/T$ diffusive behavior. With
the mixing of the YSZ component into the LCMO phase, the composite
compound, LCMO/YSZ, exhibits the usual metal-like TEP behavior in
the LT-FM phase and the usual semiconductor-like property, {\it
i.e.}, the $1/T$ dependence, in the HT-PM phase. However, in high
contrast to the result for the sample of x = 0, an abrupt jump
appears in the TEP during the phase transition for all of the
samples with x $ > $ 0 except for the one of x = 80\%. It is
difficult to make a reasonable measurement of the TEP for x = 80\%
because of its insulation property. The above results from the TEP
measurements imply that the appearance of the jump is associated
with the existence of the boundary layer between the YSZ phase
intermixed with the LCMO phase, since it appears even for the
composite ratio as little as x = 0.75 \%, considering that the YSZ
phase is insulating by itself without an experimentally-determined
TEP value and the LCMO phase alone exhibits smooth transition
during the phase transition.

By the fact that the electrical measurement does not have the
corresponding abrupt change during the phase transition, the TEP
actually reveals more subtle variation of the phase transition.
Considering the materials consisting of multiple components with
different electrical conductivity, the TEP reflects the
contributions from the separate composing ingredients weighted by
the corresponding conductivity. For the LCMO/YSZ material at x $ >
x_m $, unlike the electrical conductance which reflects
predominantly the conducting property of the components with high
conductivity, the TEP has a non-negligible or sometimes very large
contribution from the poor conducting parts of the sample because
of the fact that materials of poor conductivity usually have
higher TEP value. At x $ < x_m$, the resistivity of the LCMO phase
enclosed by the YSZ layer mainly reflects the intergranular
effect. However, for the TEP, phonon motion is the major factor to
affect the electron movement. The temperature drop across every
composition of the sample is homogeneous. Hence, it is more
sensitive to the property variation occurring in the boundary
layer. In Fig.~\ref{Tpeak}, the peak temperatures of the MI
transition, $T_{\rho}$, and the TEP maxima, $T_{S}$, are plotted
against x. The effect of the YSZ composition on these two
characteristic temperatures are similar with the minimum
temperatures occurring at $x_m$ = 4.5 \%. However, the variation
amplitude of $T_{S}$ is much less than that of $T_{\rho}$. Since
the electrical conductance is affected mainly by the high
conducting composition of the sample and the TEP, which is the
transport property of zero current, reflects the overall property
of the sample. Therefore, the TEP has smaller x dependence than
the electrical conductivity does.

Similar positive jump has been observed on the Yttrium-doped
manganites, (La$_{1-x}$Y$_{x}$)$_{2/3}$Ca$_{1/3}$MnO$_{3}$ in the
previous experiment\cite{Liu03} and has been interpreted as due to
the magnetic structural inhomogeneity caused by the doping-induced
lattice mismatch. Hence, it is reasonable to infer that the
positive jump occurring under the current investigation for the
LCMO/YSZ samples is caused by the magnetic inhomogeneity resulting
from the lattice mismatch, which is induced by the strain from the
boundary layer of the YSZ and the LCMO phases. The lattice
variation would modify the coupling strength between the adjacent
magnetic moments and result in the inhomogeneous distributions of
the PM/FM magnetic phases in the surrounding regions of the
boundary layers.

The magnetic transition is studied by the AC magnetic
susceptibility measurements. The applied AC field for the
measurements is 10 Oe at 133 Hz with a background field close to
zero. Temperature range for the measurements is from 300 K down to
10 K. The transition for all of the samples appears to be smooth.
One can determine the magnetic transition width in temperature,
$\Delta T$, by calculating d$\chi$/d$T$. For a ferromagnetic
transition, d$\chi$/d$T$ shows a peak at the transition
temperature $T_C$ with the width characterizing the transition
sharpness. A sharp transition in temperature is obtained for the
sample of x = 0 while a much wider transition, at least twice that
of the x = 0, for the rest ones. This indicates that the effect of
the magnetic inhomogeneity induced by the strain of the boundary
layer is non-negligible. In Fig.\ref{Mag}, the TEP peak value,
$S_{peak}$, and the magnetic transition width in temperature,
$\Delta T$, are plotted against x. Similar x-dependence of these
two quantities revealed in the figure further supports our
conjecture that the abrupt jump in the TEP measurement is strongly
correlated with the magnetic inhomogeneity caused by the boundary
strain.

In some of the previous experiments, magneto-thermoelectric effect
has been performed on the $La_{0.67}Ca_{0.33}MnO_3$ thin film,
\cite{Jaime96,Jaime99}. Two implications are obtained in
consistent with the explanations for our experiment. First of all,
the TEP peak appears in the thin film sample of single crystal
LCMO phase without any doping or intermixing effect. This is
different from the TEP behavior of the pure LCMO sample, x = 0,
observed in the current experiment. However, if one takes into
account the strain effect of the boundary layer adjacent to the
substrate for the thin film sample, the underlying cause for the
TEP peak would be the same as in our experiment. The second
implication is that the cause for the TEP peak is magnetic, since
the applied field would suppress the magnitude of the peak and
display the magneto-thermoelectric effect.

\section{Conclusion}

For the (1-x)La$_{2/3}$Ca$_{1/3}$MnO$_{3}$/xYSZ compound, the
existence of the YSZ insulator introduces two major effects into
the original pure LCMO materials. First of all, this is to produce
artificially the two-component coexistence structure for the
investigation of the electrical conductance within the LCMO phase.
The conductance changes from the tunnelling region at low x, with
the thin layer of YSZ insulator encompassing the LCMO phase, to
the percolation conductance region at high x, in which the YSZ
forms cluster-like structure with the LCMO phase existing in
percolation-typed connection. The peak temperature, $T_{\rho}$, in
the $\rho-T$ measurement, has a minimum at x = 4.5 \%, with the
corresponding resistivity value, $\rho_{peak}$, reaching the
maximum, inset of Fig.~\ref{RT-peak}. This corresponds to the
cross-over region from the tunnelling to the percolation-typed
conductance. The temperature of the peak in the TEP measurement,
$T_S$, has a similar x-dependence as the MI transition, but with a
much smaller variation range as shown in Fig.~\ref{Tpeak}. This is
owing to the fact it is a zero current transport property. The
poor conducting part of the sample, though contributes much less
in the electrical conductance, has an important contribution to
the TEP behavior. Secondly, the existence of the YSZ composition
introduces boundary layer into the LCMO phase. The appearance of
the TEP peak depends crucially on the existence of the YSZ
composition. It takes the YSZ ratio as little as x = 0.75\% to
induce large TEP peak, see Fig.~\ref{TEP}. This indicates that the
underlying cause for the TEP peak to occur is on the boundary
layer rather than on the bulk YSZ composition. One of the
plausible conjectures is that the strain in the boundary layer
would modify the coupling strength of the magnetic moments within
the layer and affect the conduction property due to the
inhomogeneity of the magnetic composition. From the x dependence
of the magnetic transition width in temperature, $\Delta T$,
determined by $d\chi / dT$, see Fig.\ref{Mag}, it shows that the
transition width is highly enlarged in accordance with the
existence of the YSZ phase. For the sample of x = 0, on the other
hand, the transition is much sharper. This indicates that the
boundary layer plays an important role on the magnetic structure
along with the TEP behavior during the CMR phase transition. From
the magneto-thermoelectric effect performed on the
La$_{0.67}$Ca$_{0.33}$MnO$_3$ thin film in the previous
experiments \cite{Jaime96,Jaime99}, one can reach the conclusion
that the TEP peak is induced by the lattice strain in the boundary
layer of the LCMO phase and is magnetic in origin.

\section{Acknowledgement}

This work is partially supported by the Doctoral Foundation of the
National Ministry of Education through the Grant No. 2000000146 of
the P.\,R.\,China.

\begin{figure}
\caption{The $\rho-T$ curves for the LCMO/YSZ samples with x = 0,
0.75\%, 1.25\%, 4.5\%, 13\%, and 15\%. The inset presents the x
dependence of the peak temperature, $T_P$, and the resistivity of
the peak, $\rho_{peak}$, in the $\rho-T$ measurements. }
\label{RT-peak}
\end{figure}

\begin{figure}
\caption{The temperature dependence of TEP from 77 K to 320 K for
the samples of x =  0, 0.75\%, 1.25\%, 4.5\%, 13\%, and 15\%. The
sample of x = 0 is the pure LCMO polycrystalline.} \label{TEP}
\end{figure}

\begin{figure}
\caption{The x dependence of the peak temperatures, $T_P$, in the
$\rho-T$ measurement and, $T_S$, in the TEP measurement. }
\label{Tpeak}
\end{figure}

\begin{figure}
\caption{The x dependence of the TEP peak value, $S_{peak}$, and
the magnetic transition width in temperature, $\Delta T$.  }
\label{Mag}
\end{figure}

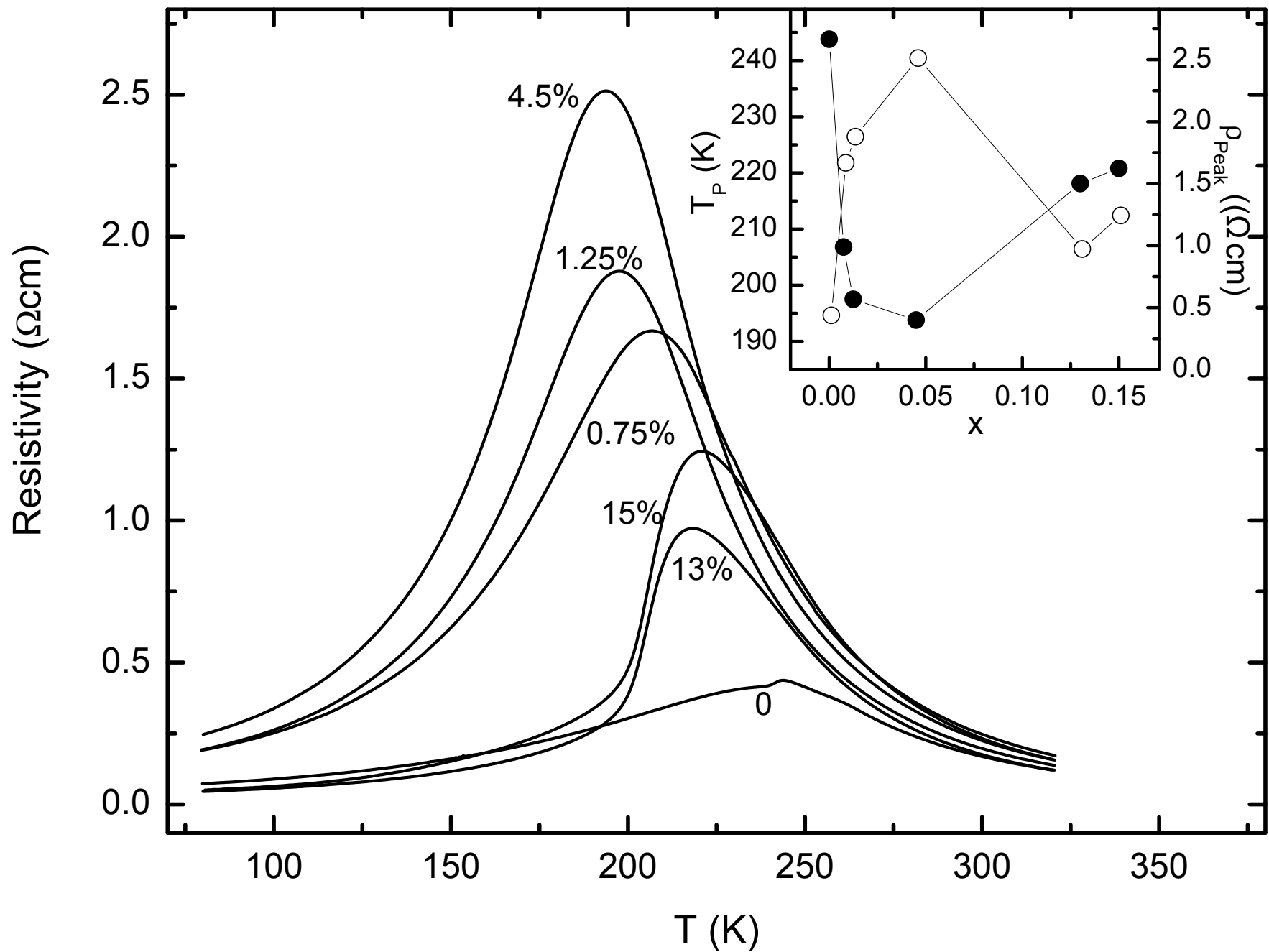
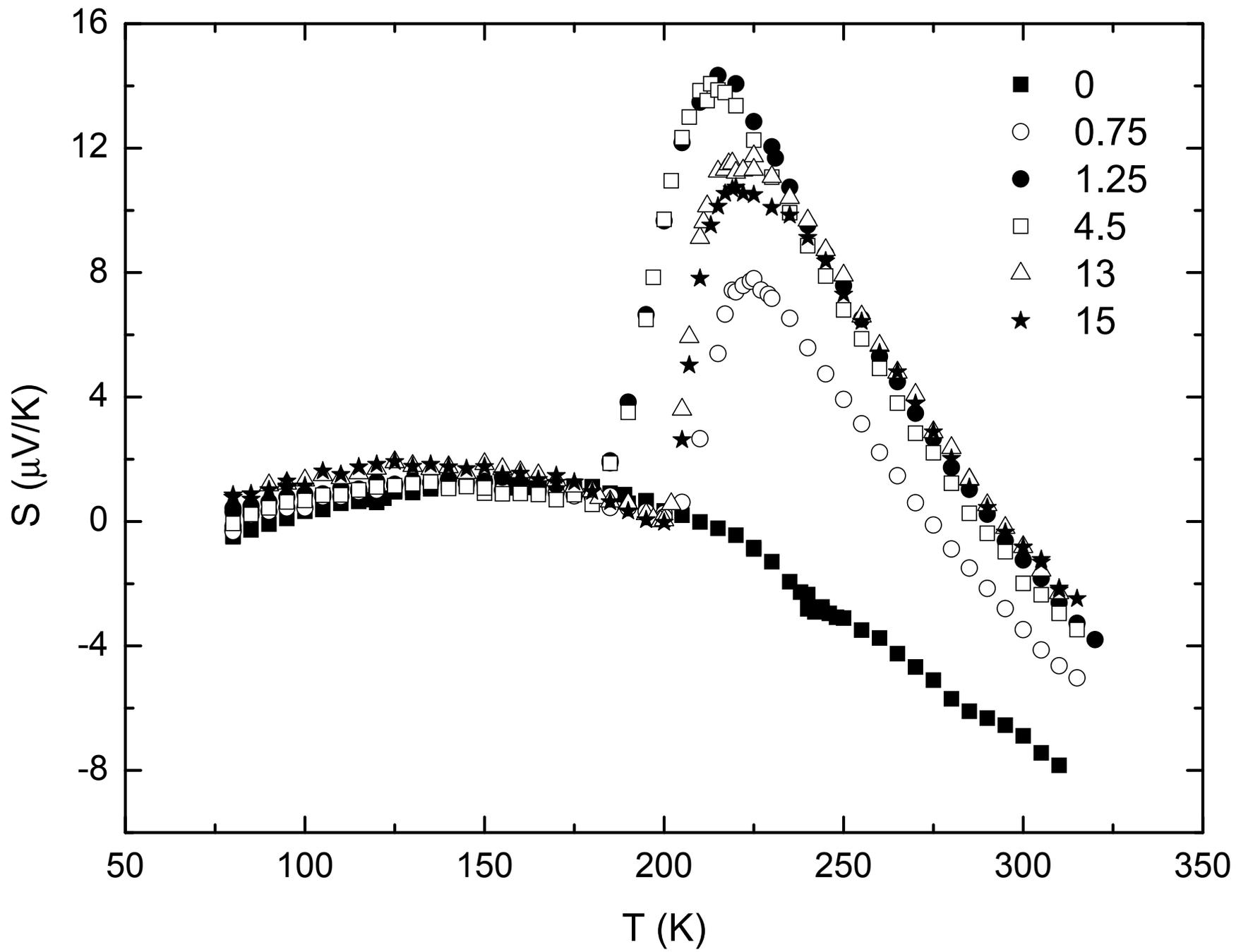
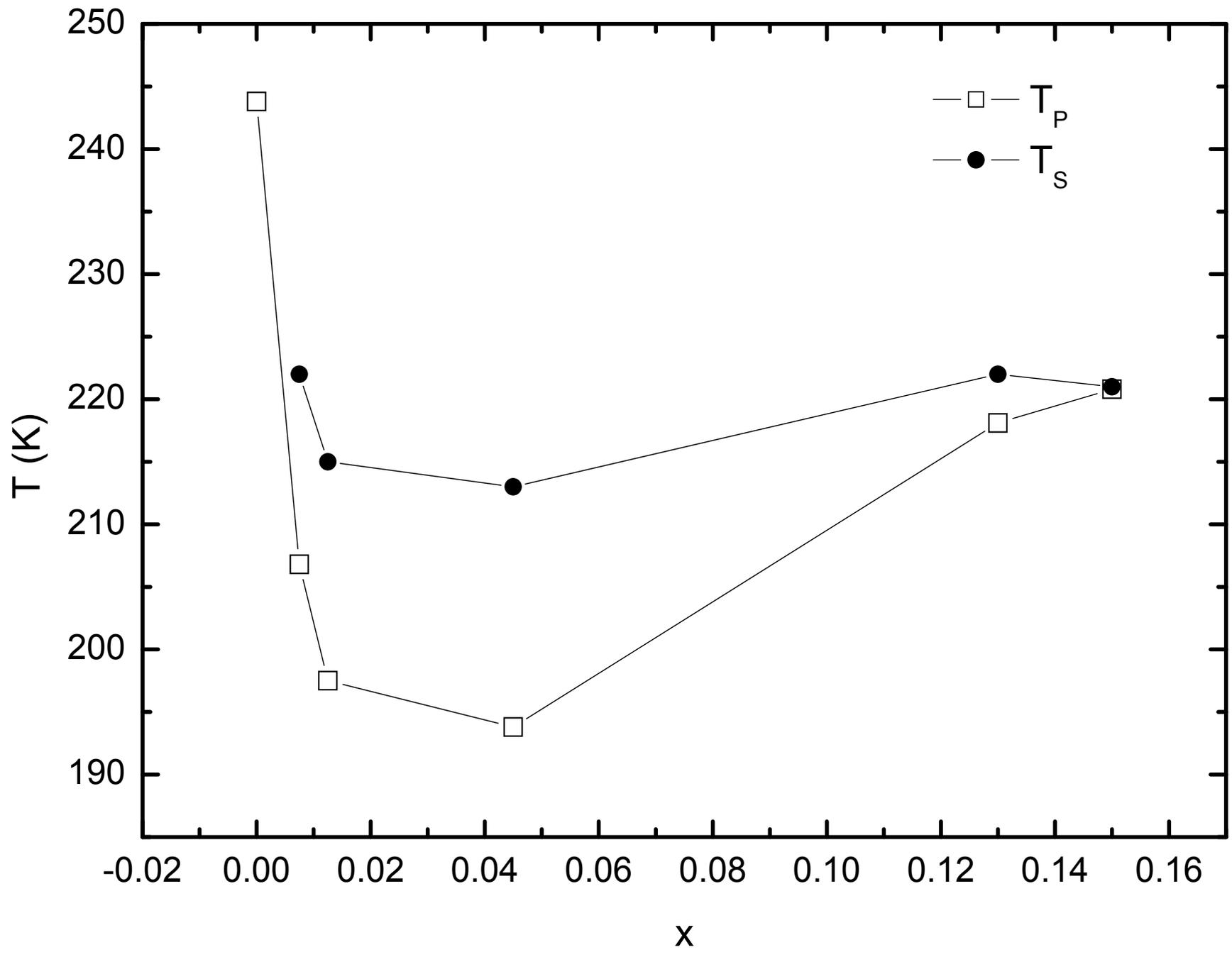
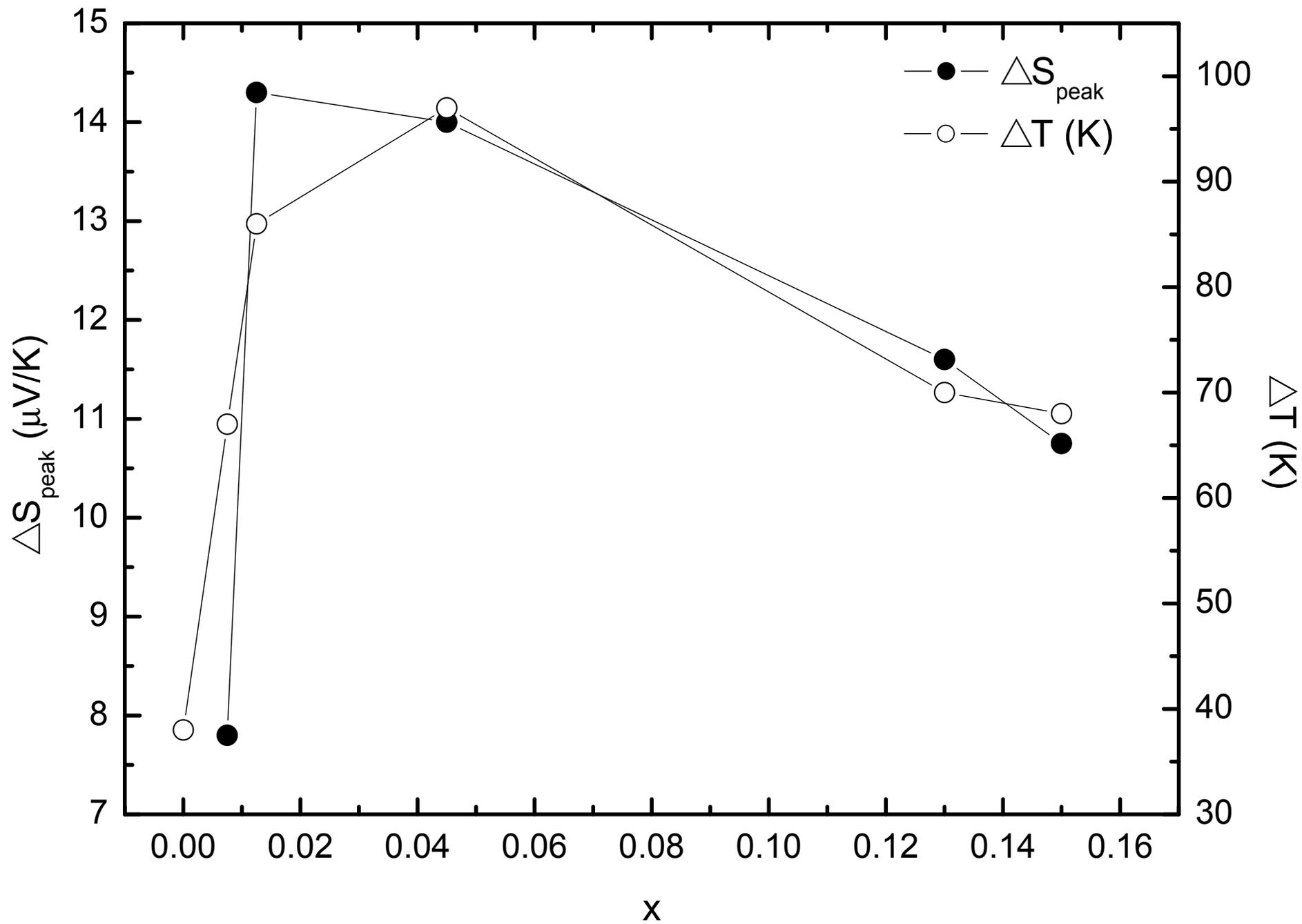
\end{document}